\documentclass[10pt]{article}
\usepackage{graphicx}

\def\Title#1{\begin{center} {\Large #1 } \end{center}}
\def\Author#1{\begin{center}{ \sc #1} \end{center}}
\def\Address#1{\begin{center}{ \it #1} \end{center}}

\newcommand\pubblock{\rightline{\begin{tabular}{l} Proceedings of the Second Annual LHCP\\ \pubnumber\\
         \pubdate  \end{tabular}}}

\newenvironment{Abstract}{\begin{quotation} \begin{center} 
             \large ABSTRACT \end{center}\bigskip 
      \begin{center}\begin{large}}{\end{large}\end{center} \end{quotation}}

\newenvironment{Presented}{\begin{quotation} \begin{center} 
             PRESENTED AT\end{center}\bigskip 
      \begin{center}\begin{large}}{\end{large}\end{center} \end{quotation}}

\def\beq{\begin{equation}}
\def\eeq#1{\label{#1}\end{equation}}
\def\eeqn{\end{equation}}

\def\beqa{\begin{eqnarray}}
\def\eeqa#1{\label{#1}\end{eqnarray}}
\def\eeqan{\end{eqnarray}}

\def\overbar#1{\overline{#1}}

\let\bar=\overbar

\def\L{{\cal L}}

\def\Dslash{\not{\hbox{\kern-4pt $D$}}}
\def\dslash{\not{\hbox{\kern-2pt $\del$}}}

\def\msb{{\bar{\ssstyle M \kern -1pt S}}}

\usepackage{color}

\newcommand\pubnumber{ ATL-UPGRADE-PROC-2014-004 }

\newcommand\pubdate{\today}

\def\affiliation{
On behalf of the ATLAS Collaboration, \\
Istituto Nazionale di Fisica Nucleare,  Genova, Italy }

\begin{document}

\large
\begin{titlepage}
\pubblock

\vfill
\Title{  The ATLAS upgrade program  }
\vfill

\Author{ Claudia Gemme  }
\Address{\affiliation}
\vfill
\begin{Abstract}

  After the first successful LHC run in 2010-2012, plans are actively advancing for a series of
upgrades 
leading eventually to annual integrated luminosities about ten times the current design after 10 years. 
The larger integrated luminosity will allow to perform precise measurements of the just discovered Higgs boson
and to continue searching for new physics beyond the Standard Model. 
 Coping with the high instantaneous and integrated luminosity will be a great  challenge for the ATLAS detector and
will require  changes in most of the subsystems, especially those at low radii
or large pseudorapidity, as well as in its trigger architecture. 
Plans to consolidate and, whenever possible, to improve the physics performance of the current detector over the next decade are summarized in this paper.

\end{Abstract}
\vfill

\begin{Presented}
The Second Annual Conference\\
 on Large Hadron Collider Physics \\
Columbia University, New York, U.S.A \\ 
June 2-7, 2014
\end{Presented}
\vfill
\end{titlepage}
\def\thefootnote{\fnsymbol{footnote}}
\setcounter{footnote}{0}
%

\normalsize 


\section{Introduction}

ATLAS~\cite{ATLAS_Jinst}
is a general-purpose experiment designed to explore the
$pp$ collisions  at the CERN Large Hadron Collider (LHC)~\cite{LHC_Jinst} at center of mass
energies  up to $\sqrt{s}$ = 14 TeV  and a maximum peak luminosity 
of 10$^{34}$ cm$^{-2}$s$^{-1}$.

ATLAS has been successfully operating in 2010-2012, collecting collision data 
at $\sqrt{s}$ = 7 and 8 TeV, recording an integrated luminosity of 
$\sim$ 25 fb$^{-1}$ with a peak luminosity of
$7.7 \times $10$^{33}$ cm$^{-2}$s$^{-1}$.  
The discovery \cite{Higgs_ATLAS, Higgs_CMS} of a 126 GeV Higgs boson in 2012 is  the greatest, even if not the unique, achievement 
allowing to identify the last piece of the highly successful Standard Model (SM).
In the next years, LHC will undergo a series of upgrades
leading ultimately to five-fold increase of the instantaneous
luminosity with leveling according to the High-Luminosity LHC (HL-LHC)
project. The goal is to extend the dataset from about
300 fb$^{-1}$, expected to be collected by the end of the LHC
run (in 2022), to 3000 fb$^{-1}$ by 2035. The foreseen
higher luminosity at the HL-LHC is a great challenge for
ATLAS. Meeting it will require significant but gradual detector optimizations,
changes and improvements, which are the subject
of these proceedings.

\section{LHC and ATLAS Upgrade Plans}

\begin{figure}[htb]
\centering
\includegraphics[width=0.35\textwidth]{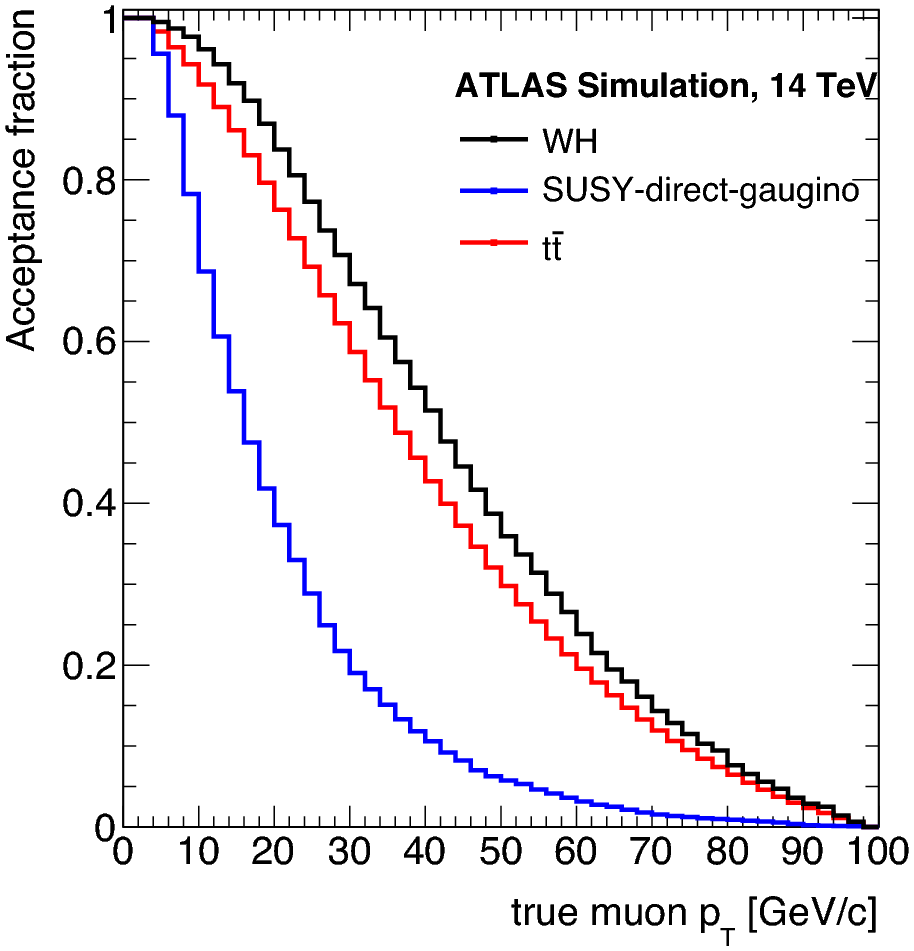}
   \vspace{-0.4cm}
\caption{Acceptance of muons from $t\overbar{t}$, WH and SUSY processes as a function of true muon momentum.}
\label{fig:trigger}
\end{figure}

The main motivation for the LHC upgrades is to extend and improve the physics program.
The capability to maintain an optimal trigger system as the luminosity increases
beyond its nominal design value is crucial for the physics reaches  and requires a strong reduction of the main source of backgrounds,
 as jets mimicking electrons in the calorimeters and misidentified muons in the forward spectrometer.
Otherwise, increased trigger threshold cuts would have to be deployed to control the rates, reducing
significantly the signal efficiency. As an example, the acceptance for muons from $t\overbar{t}$, WH and 
SUSY processes as a function of the true muon $p_{T}$ is shown in Figure~\ref{fig:trigger}, where an increase in
threshold from 20 GeV to 30 GeV results in a reduction in signal acceptance ranging from 1.3 to 1.8.

\begin{figure}[htb]
\begin{center}
\begin{tabular}{cc}
     \includegraphics[width=0.4\textwidth]{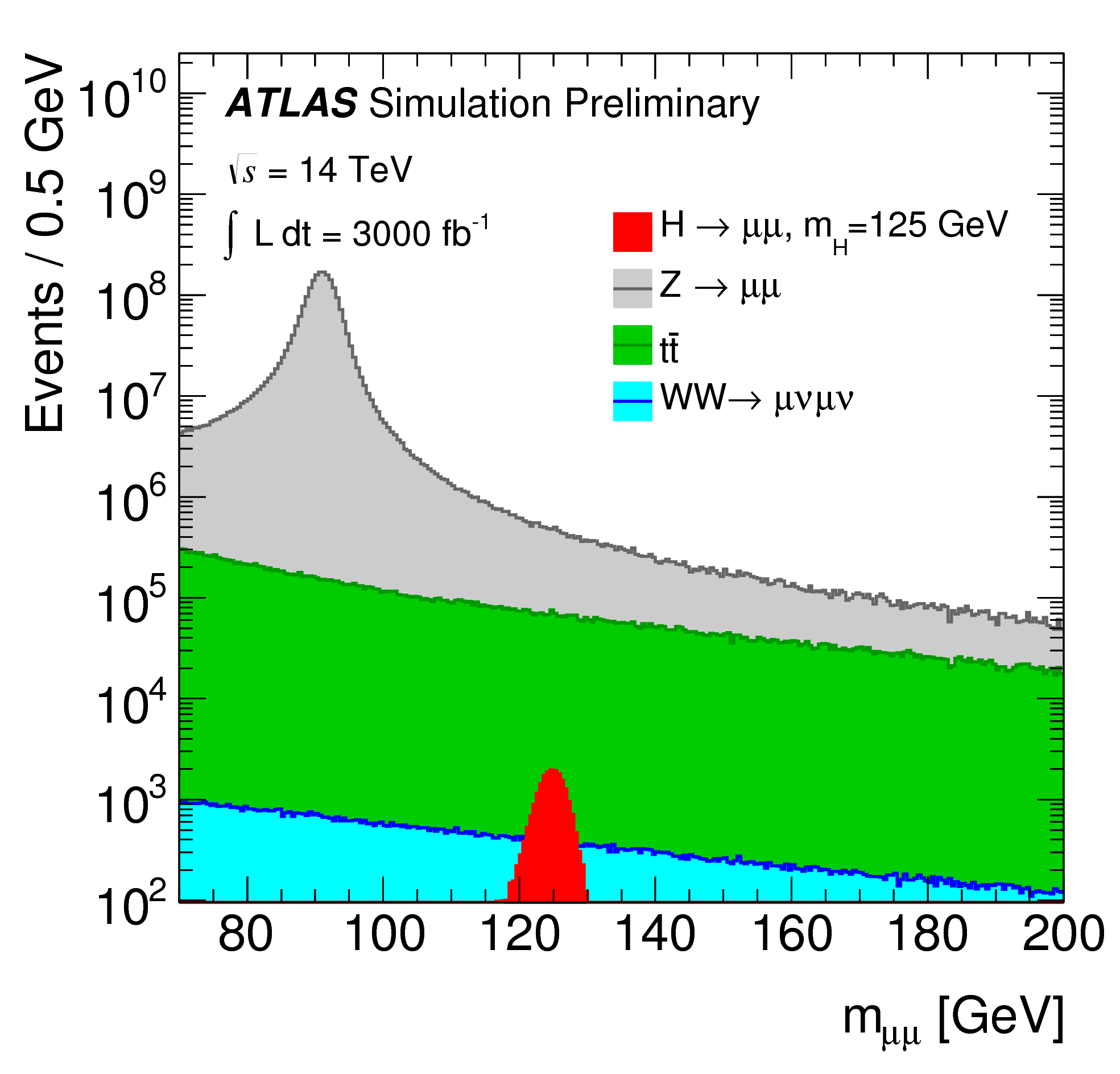} &
   \hspace{0.5cm}
     \includegraphics[width=0.24\textwidth]{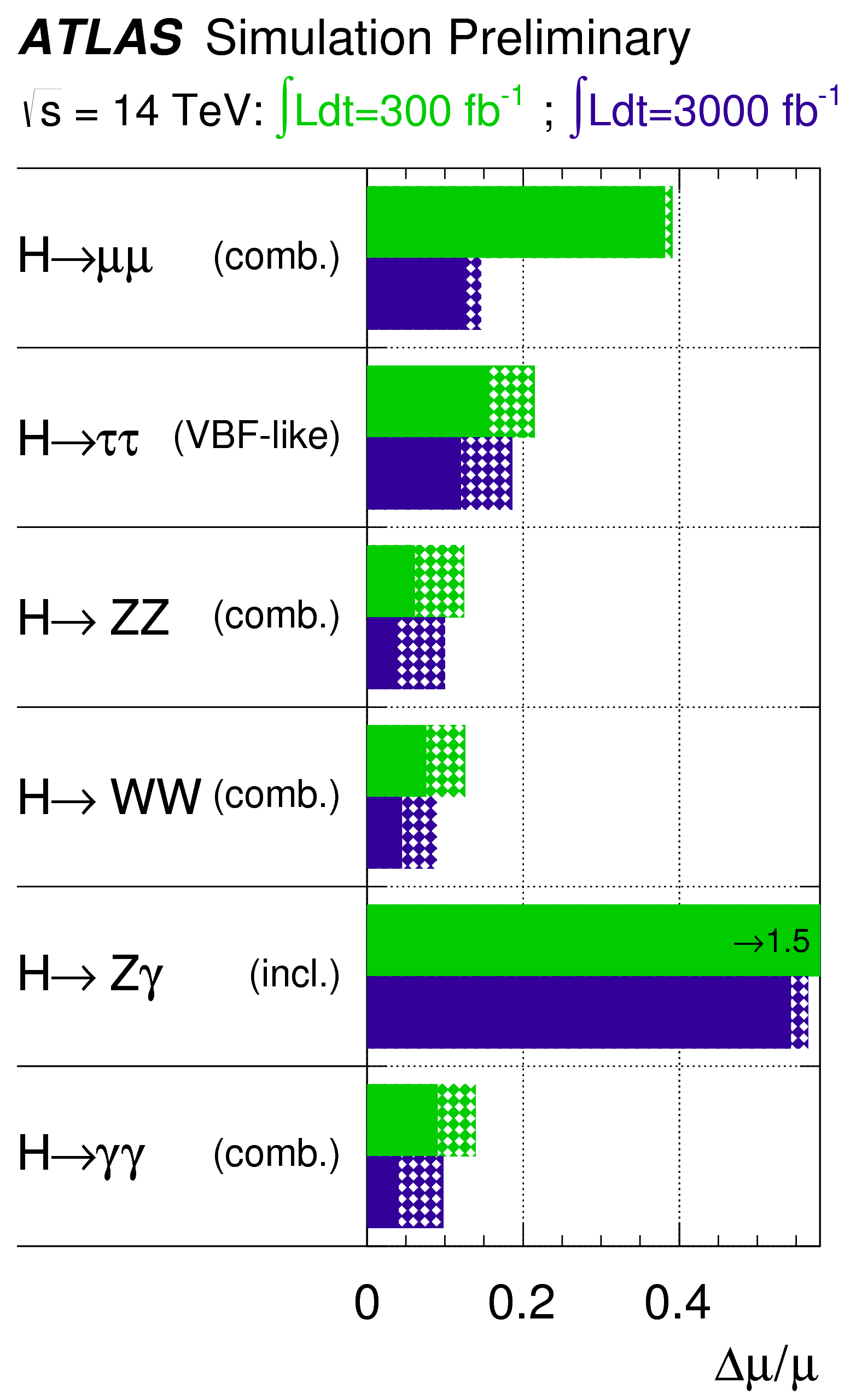} \\
\end{tabular}
   \end{center}
   \vspace{-0.5cm}
\caption{
(a) Distribution of the dimuon invariant mass of the signal and background processes generated for $\sqrt{s} = 14$ TeV and $\L$ = 3000 fb$^{-1}$. 
(b) Relative uncertainty on the total signal strength for all Higgs final states in the different
experimental categories used in the combination, assuming a SM Higgs Boson with a mass of 125 GeV. The hashed areas indicate the increase of the estimated error
due to current theory systematic uncertainties.
}
\label{fig:HiggsUpgrade}
\end{figure}

A major focus of the physics program is on the Higgs boson~\cite{HiggsUpgradePubNote}. 
With larger luminosity, it will be possible to make precision measurements of the Higgs boson properties,
in particular its couplings to fermions and bosons, its rare decays and its self-couplings.  As an example, 
Fig.~\ref{fig:HiggsUpgrade}(a) shows the dimuon invariant mass for the measurement of the   H $\rightarrow \mu \mu$ channel: with an integrated
luminosity of 3000 fb$^{-1}$, this rare  channel can be observed.
The relative uncertainties on the total signal strength for some Higgs decay modes 
for the 300 and 3000 fb$^{-1}$ luminosities are  reported in  Fig.~\ref{fig:HiggsUpgrade}(b). 
New physics searches, beyond the Standard Model, will continue in SUSY or other exotic physics scenarios. 
$WW$ scattering measurements will  also be essential amongst the
electroweak symmetry breaking mechanism measurements.

The harsher radiation environment and higher detector occupancies at the HL-LHC imply major changes to
most of the ATLAS systems, specially those at low radii and large pseudorapidity. A general guideline for these
changes is maintaining the same or improve the detector performance measured at the LHC. The higher event
rates and event sizes will be a challenge for the trigger and
data acquisition (DAQ) systems, which will require a significant
extension of their capacity.

The ATLAS upgrade will be gradual and flexible to accommodate a possible evolution of LHC operational parameters  and indications of  new physics signals. 
It is planned in three phases, which correspond to the three long technical shutdowns of the
LHC towards the HL-LHC.

\section{Phase-0 upgrades}
The repair of the splices in the main accelerator during the long shutdown in 2013-2014 (LS1) will allow the LHC
to continue its operation  close to its design parameters with a  center of mass energy  $\sqrt{s}$ = 13-14 TeV, 25 ns bunch spacing    and
peak luminosities $\sim 1 \times 10^{34}$ cm$^{-2}$s$^{-1}$, and deliver an integrated luminosity  $\geq$50 fb$^{-1}$ before the following shutdown. 

\begin{figure}[htb]
\centering
\includegraphics[width=0.5\textwidth]{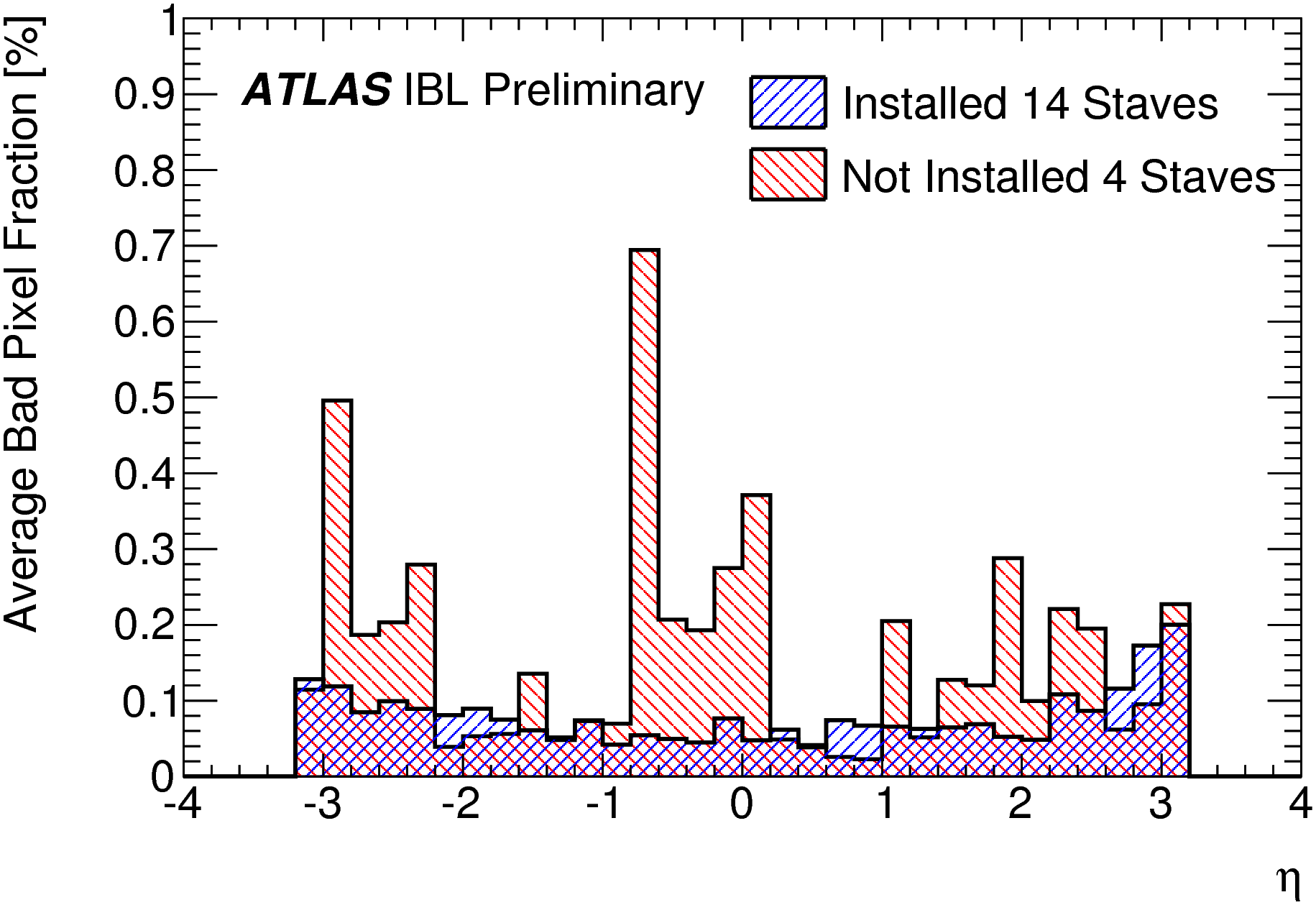}
   \vspace{-0.6cm}
\caption{ Average bad pixel fraction distribution as a function of $\eta$ for installed (blue) and not
installed (red) production staves.}
\label{fig:IBL}
\end{figure}

ATLAS has used  the shutdown period (Phase-0)  for detector consolidation works, including
 a new Inner Detector (ID) cooling system, a new diamond beam monitor, the replacement of the Pixel internal services, 
new power supplies for the calorimeter, an improved  coverage of the Muon Spectrometer (MS) between the barrel and the end-cap regions and a new beam pipe, in the central and forward region. 
But the main upgrade activity in Phase-0 is the
installation of a new barrel layer in the Pixel detector.
The Insertable B-Layer (IBL)~\cite{IBL_TDR} is an additional, 4$^{th}$ pixel
layer, that has been built around the new central beam pipe and then slipped inside the present Pixel detector in situ.
The IBL is therefore placed between the actual innermost pixel
layer (the B-layer) and the beam pipe, at a sensor average radius of  33 mm (50.5 mm is the radius of the B-layer).
  To make the installation of the IBL possible, the new Be beam pipe radius  is 
reduced by 4 mm radius (r=29 mm $\to$ r=25 mm). \\ 
IBL consists of 14 pixel staves surrounding the beam-pipe.
Each carbon-fibre stave carries and provides cooling to 32
read-out chips, which are bump-bonded to silicon sensors. 
The pseudo-rapidity coverage extends to $|\eta|<$3.
Two types of sensors are used:
 planar n-in-n sensors, similar to the present Pixel detector, and 3D silicon
sensors installed, for the first time,  in a tracking detector. To cope with a larger fluence and peak luminosity, higher
hit rate and occupancy, a new generation of read-out chip, FE-I4, has been developed using
a new architecture, IBM 130~nm CMOS process manufacturing resulting in a smaller pixel size  ($50 \times 250~\mu\rm{m}^2$). 
18 staves have been qualified and out of them the best 14 have been chosen to be assembled in the IBL detector. Fig.~\ref{fig:IBL} shows the longitudinal distribution of the bad pixel fraction for the produced staves. The new detector has been installed inside ATLAS in May 2014 and commissioning is ongoing. 
So far it is fully functional with only 0.1\% of the channels disabled. \\
 IBL will help to preserve the tracking
performance at high luminosity when the B-layer will suffer from radiation damage
and high pile-up occupancies. Moreover, it will compensate for defects (irreparable failures
of modules) in the existing detector, assuring tracking
robustness. Given the closer position to the impact point and the smaller pixel size, IBL
will also improve the vertex resolution, secondary vertex finding
and $b$-tagging, hence extending the reach of the physics
analysis. 

\section{Phase-I upgrades}
A second shutdown (LS2) is being planned in 2018 to integrate the Linac4 into the injector
complex, to increase the energy of the PS Booster to reduce the beam emittance, and to upgrade the
collider collimation system. When data taking resumes in 2020, the peak luminosity is
expected to reach $\sim 2-3 \times 10^{34}$ cm$^{-2}$s$^{-1}$ corresponding to 55 to 80 interactions per crossing (pile-up)
with 25 ns bunch spacing, well beyond the initial design goals. LHC is planning to deliver  an integrated luminosity of $\sim$ 300 fb$^{-1}$ before the next shutdown, thus 
extending the reach for discovery of new physics and the ability to study new phenomena and states. 
In this long shutdown, so-called Phase-I, ATLAS proposes the installation of new muon Small Wheels  and 
 several  updates for the trigger system (Fast TracKer, Calorimetric and topological triggers)  to handle luminosities well beyond the nominal values. 
Detailed plans are described in Ref.~\cite{LoI-PhaseI}.

\begin{figure}[thb]
\begin{center}
\begin{tabular}{cc}
     \includegraphics[width=0.375\textwidth]{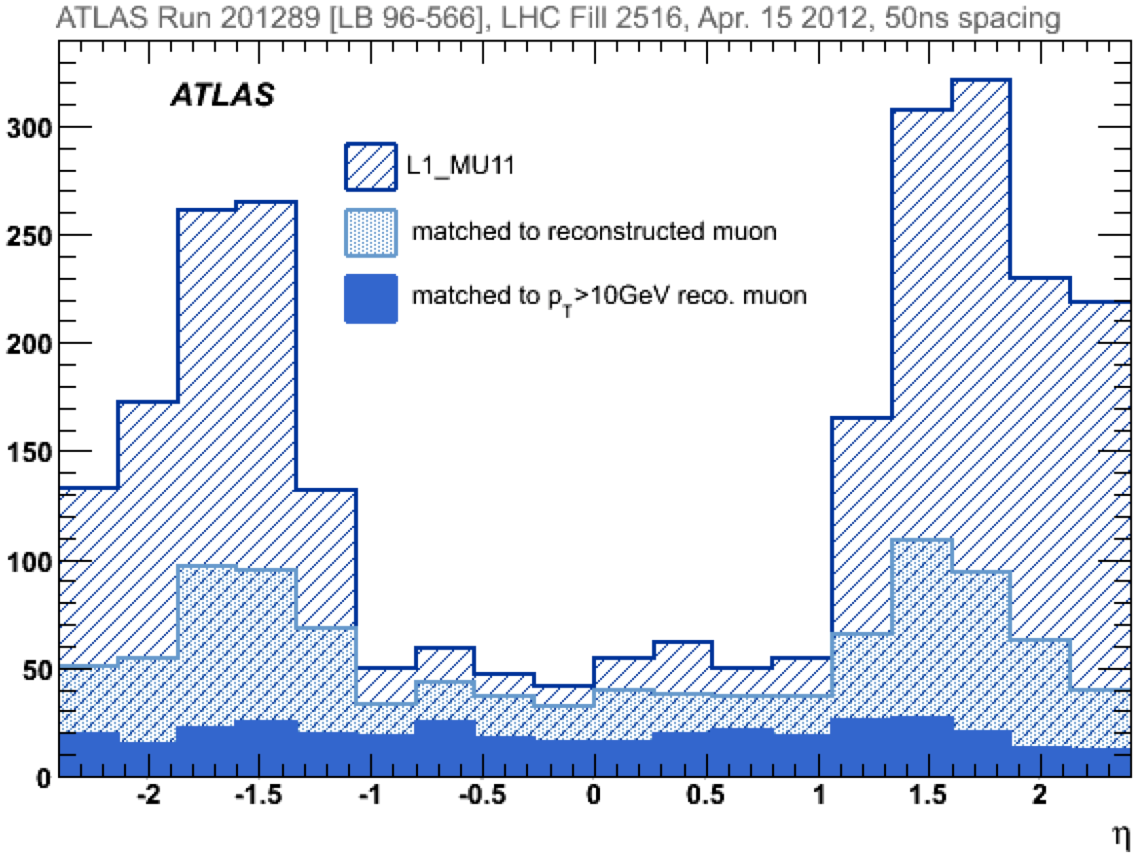} &
   \hspace{0.3cm}
     \includegraphics[width=0.43\textwidth]{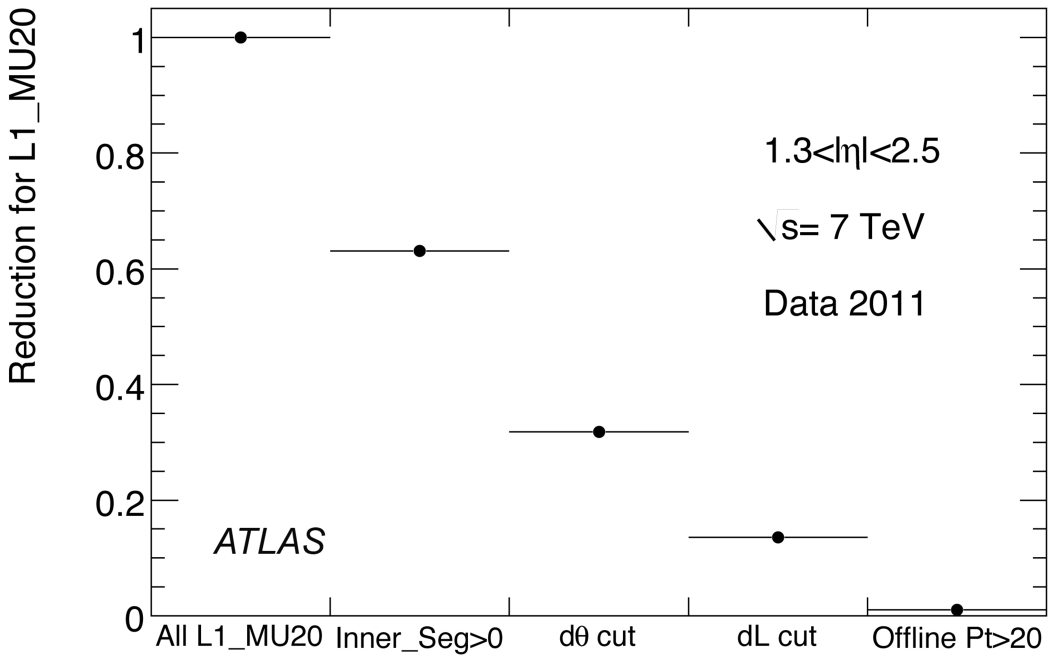} \\
\end{tabular}
   \end{center}
   \vspace{-0.35cm}
\caption{
$\eta$ distribution of Level-1 muon signal L1\_MU11 with the distribution of the subset with matched muon candidate to an offline well reconstructed muon and offline
reconstructed muons with $p_T > $10 GeV. 
(b) Expected reduction of L1\_MU20 rate simulated using collision data by successive applications
of requirements on the small wheel segments: to have a segment in the Small Wheel 
(Inner\_Seg $>$ 0), that it is pointing to the primary
 (d$\theta$ cut) and matched to the external muon chamber segment (dL cut).}
\label{fig:Muons}
\end{figure}

 {\sc\emph{ \textbf{New Small Wheels}}}.
At high luminosity the performance of the muon tracking chambers (in particular in the end-cap region)
degrades with the expected increase of cavern background rate. Moreover, in the current detector,
the Level-1 muon trigger in the end-cap region is based on track segments in the intermediate muon station 
 located after the end-cap toroid magnet. An analysis of 2012 data demonstrates, see Fig.~\ref{fig:Muons}(a), 
that approximately 90\% of
the muon triggers in the end-caps are fake or background, dominated by low energy particles, mainly
protons, generated in the material located between the Small Wheel and the middle station,
hitting the end-cap trigger chambers at an angle similar to that of real high $p_T$ muons.
 Therefore, to  face with the large expected rates of the LHC upgrades,  
a replacement of the first endcap station of the Muon Spectrometer,
the New Small Wheel (NSW), is  proposed~\cite{NSW_TDR}. 
The NSW must ensure efficient tracking
at high particle rate (up to $5 \times 10^{34}$ cm$^{-2}$s$^{-1}$)
and larger $|\eta|$, with position resolution of $< 100~\mu$m. Furthermore, unlike the present layer,
the NSW will be integrated into the Level-1 trigger, thus helping in rejecting background by selecting  tracks coming from the primary interaction and 
matched with the most external layers of the muon spectrometer, see Fig.~\ref{fig:Muons}(b).

The NSW will have two chamber
technologies, one primarily devoted to the Level-1 trigger function (small-strip Thin Gap Chambers,
sTGC) and one dedicated to precision tracking (MicroMegas detectors, MM). The sTGC are
primarily deployed for triggering given their single bunch crossing identification capability. The
MM detectors have exceptional precision tracking capabilities due to their small gap (5 mm) and
strip pitch (approximately 0.5 mm). Such a precision is crucial to maintain the current ATLAS
muon momentum resolution in the high background environment of the upgraded LHC.

\begin{figure}[htb]
\begin{center}
\begin{tabular}{cc}

     \includegraphics[width=0.43\textwidth]{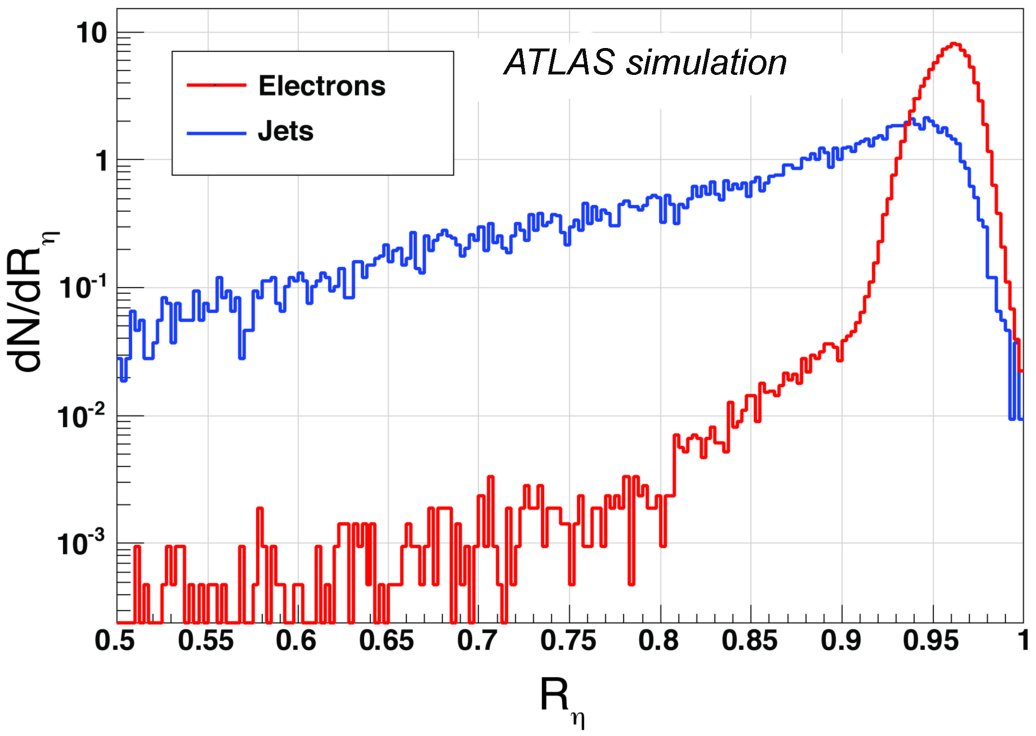} &
   \hspace{0.3cm}
     \includegraphics[width=0.43\textwidth]{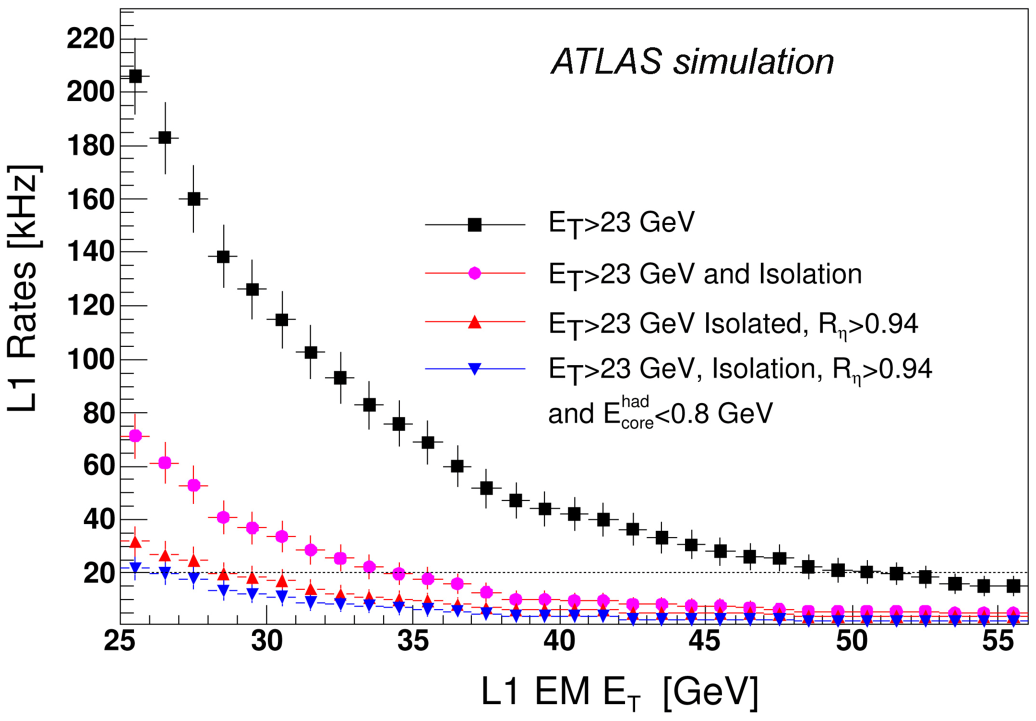} \\
\end{tabular}
   \end{center}
   \vspace{-0.5cm}
\caption{Monte Carlo simulations: (a) Distribution of the R$_{\eta}$ parameter for electrons and jets, defined as the ratio of the energy in the 3x2 over the energy in the 7x2 clusters of the 2nd layer of the EM calorimeter. (b) Expected Level-1 rates for different algorithms and conditions: the current Level-1 rate for non-isolated EM objects, for isolated EM objects, for isolated EM objects after a shower shape R$_{\eta} > $0.94 cut applied.}
\label{fig:Calo}
\end{figure}

 {\sc\emph{ \textbf{Level-1 LAr Calorimeter Electronics}}}.
Also the calorimetric trigger will have an upgrade~\cite{LAR_TDR} in Phase-1: 
the objective of this upgrade is to provide higher-granularity, higher-resolution and longitudinal
shower information from the calorimeter to the Level-1 trigger processors. The 10-fold increase
in granularity will  improve
the trigger energy resolution and efficiency for selecting electrons, photons, $\tau$ leptons, jets and
missing transverse momentum, while enhancing discrimination against backgrounds and
fakes. Fig.~\ref{fig:Calo} shows how the finer granularity enables a more sophisticated
rejection of jet backgrounds than in the current system through the use of shower shape
variables: in the (a) distribution the discriminating power of one variable is shwon, while (b) shows the Level-1 trigger rate reduction when exploiting it. 

 {\sc\emph{ \textbf{Fast Track Trigger (FTK)}}}.
The Fast TracKer Trigger~\cite{FTK_TDR} will perform the track finding and fitting  on-line using dedicated  massive parallel processing, which makes it extremely faster, instead of the Level-2 software farm as in the current trigger schema. FTK will then provide the track parameters with resolution close to the offline one shortly after the start of the Level-2 processing thus releasing  extra resources
for more advanced selection algorithms, which ultimately could improve the performances of the tracking-based filter algorithms such as the $b$-tagging and $\tau$ trigger. While the full geometrical coverage for full Phase-I pile-up  in foreseen after the 2018 shutdown, a progressive coverage and commissioning will start already in 2015.

 {\sc\emph{ \textbf{Trigger and Data Acquisition}}}.
New calorimeter feature extraction processors and a new processor for the muon signals will produce object quantities that can be combined with each other and other signals in a new topological processor. The Data Acquisition and the High-Level Trigger (HLT) processing farm will be upgraded to allow full calorimetry information
and the outputs of the FTK system to be read out and processed. A detailed description of this upgrade can be found in \cite{TDAQ_TDR}.

\section{Phase-II upgrades}
The Phase-I upgrades are designed to be fully forward-compatible with the physics program of the
high luminosity HL-LHC (Phase-II), when the instantaneous luminosity should reach $\sim 5-7 \times 10^{34}$ cm$^{-2}$s$^{-1}$, up to 200 interactions per crossing (pile-up) and
a total integrated luminosity of 3000 fb$^{-1}$. A third 36-months long shutdown (LS3) in  2023-25 will be necessary to upgrade the accelerator to this ultimate  operation mode. ATLAS is being planning major updates in all its subsystems  and trigger architecture \cite{LoI-PhaseII} .

\begin{figure}[htb]
\begin{center}
\begin{tabular}{cc}
     \includegraphics[width=0.43\textwidth]{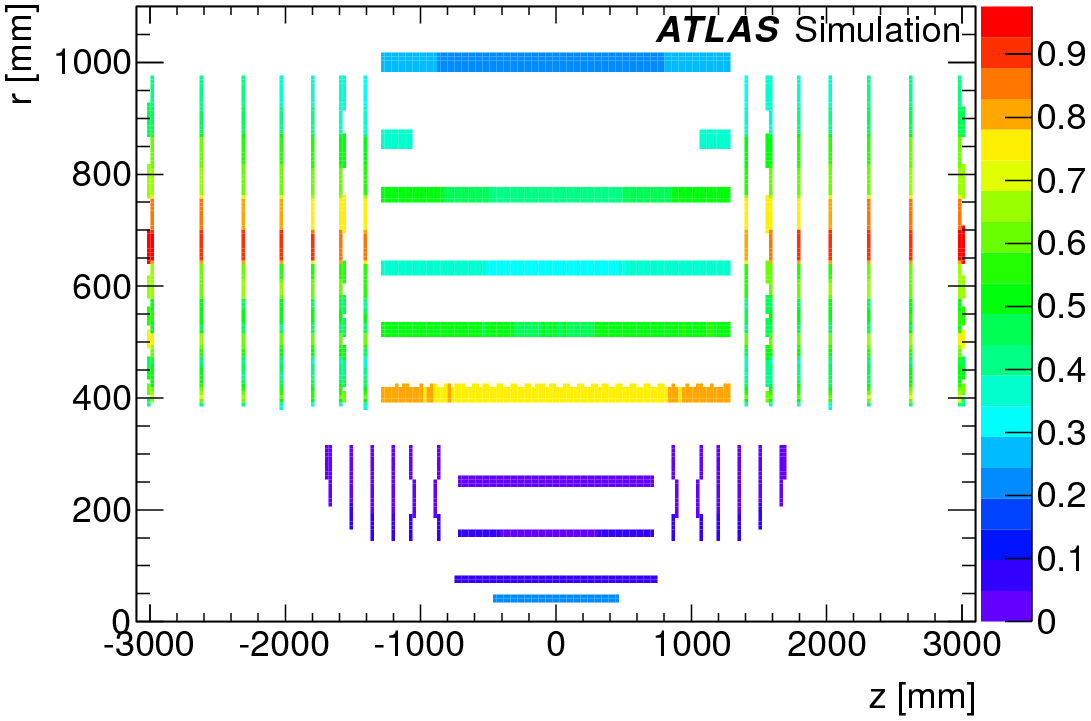} &
   \hspace{0.3cm}
     \includegraphics[width=0.395\textwidth]{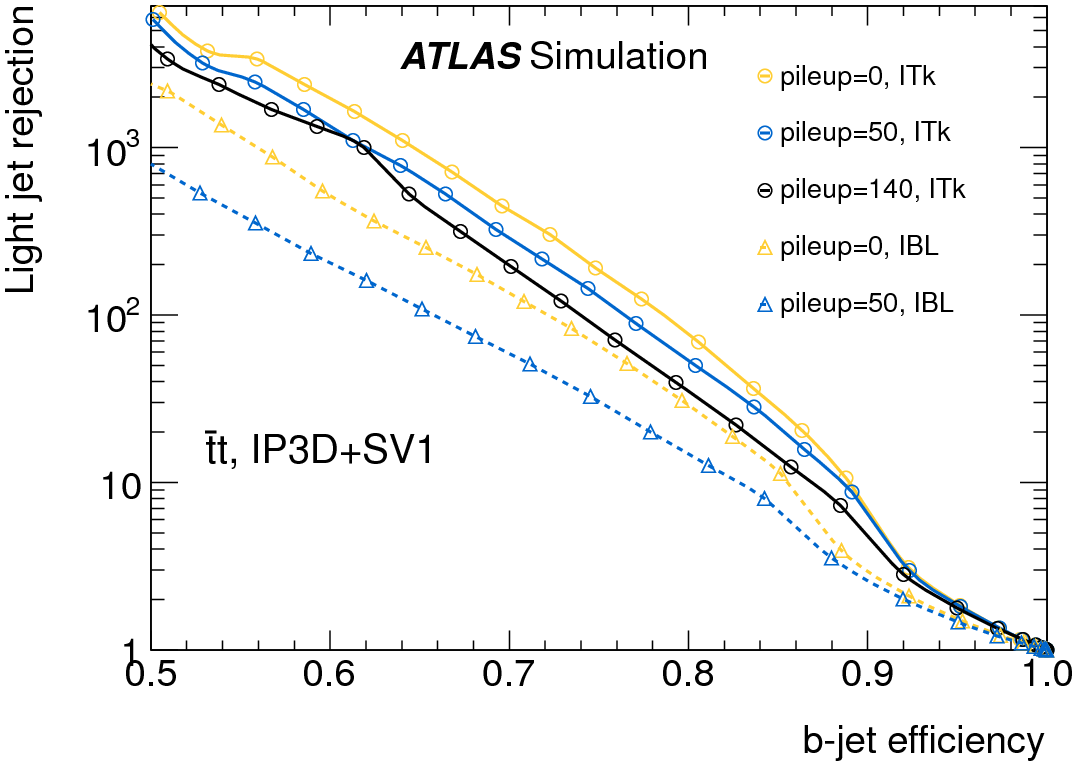} \\
\end{tabular}
   \end{center}
   \vspace{-0.5cm}
\caption{(a) Channel occupancies (in percent) with 200 pile-up events. (b) Performance of $b$-tagging in $t\overbar{t}$  events, for a range of pile-up levels for the proposed Phase-II
Tracker layout in comparison with ID+IBL.}
\label{fig:ITK}
\end{figure}

 {\sc\emph{ \textbf{New Inner Tracker}}}.
The present ATLAS Inner tracker will have several limitations in Phase-II when up to 200 pile-up events per bunch crossing are expected.
The gas-based TRT (Transition Radiation Tracker, the outermost system of the the ATLAS Inner Detector) has a limit due  to  instantaneous luminosity because of very high occupancy.  The functionality of the silicon-based
parts of the tracker  will be deteriorated due to the  total radiation dose affecting both sensors and read-out electronics and
also by the instantaneous luminosity, too high for the present limited band-width.
Because of all these factors, the entire Inner Detector will be replaced with a new, all-silicon Inner
Tracker (ITk) with pixel sensors at the inner radii surrounded
by microstrip sensors. The current baseline design of the ITk, with a  layout similar to the present detector, 
has in the central region sensors arranged in cylinders, with
4 pixel layers followed by 3 short-strip layers then 2 long-strip layers;
the forward regions will be covered by 6 pixel disks and 7 strip disks. Other layouts are currently under study, even extending
the coverage at larger pseudo-rapidity. 
Some characteristics of the performance of this layout are listed: robust tracking with at least 11 hits/track for $|\eta| < 2.5$, 
channel occupancy $< 1\%$ for pile-up up to 200 (Fig.~\ref{fig:ITK}(a)), reduced material (factor 5 for  $|\eta| < 1$)
with respect to current ID.
The performance of $b$-tagging, in particular for high energy jets, depends critically on the
rate of fake tracks, and on the two-track resolution. Using current algorithms,
the light jet rejection of the new tracker at high pile-up is found (Fig.~\ref{fig:ITK}(b)) to be as
good as that for the current ID (including the IBL) at zero pile-up.

\begin{figure}[htb]
\begin{center}
     \includegraphics[width=0.7\textwidth]{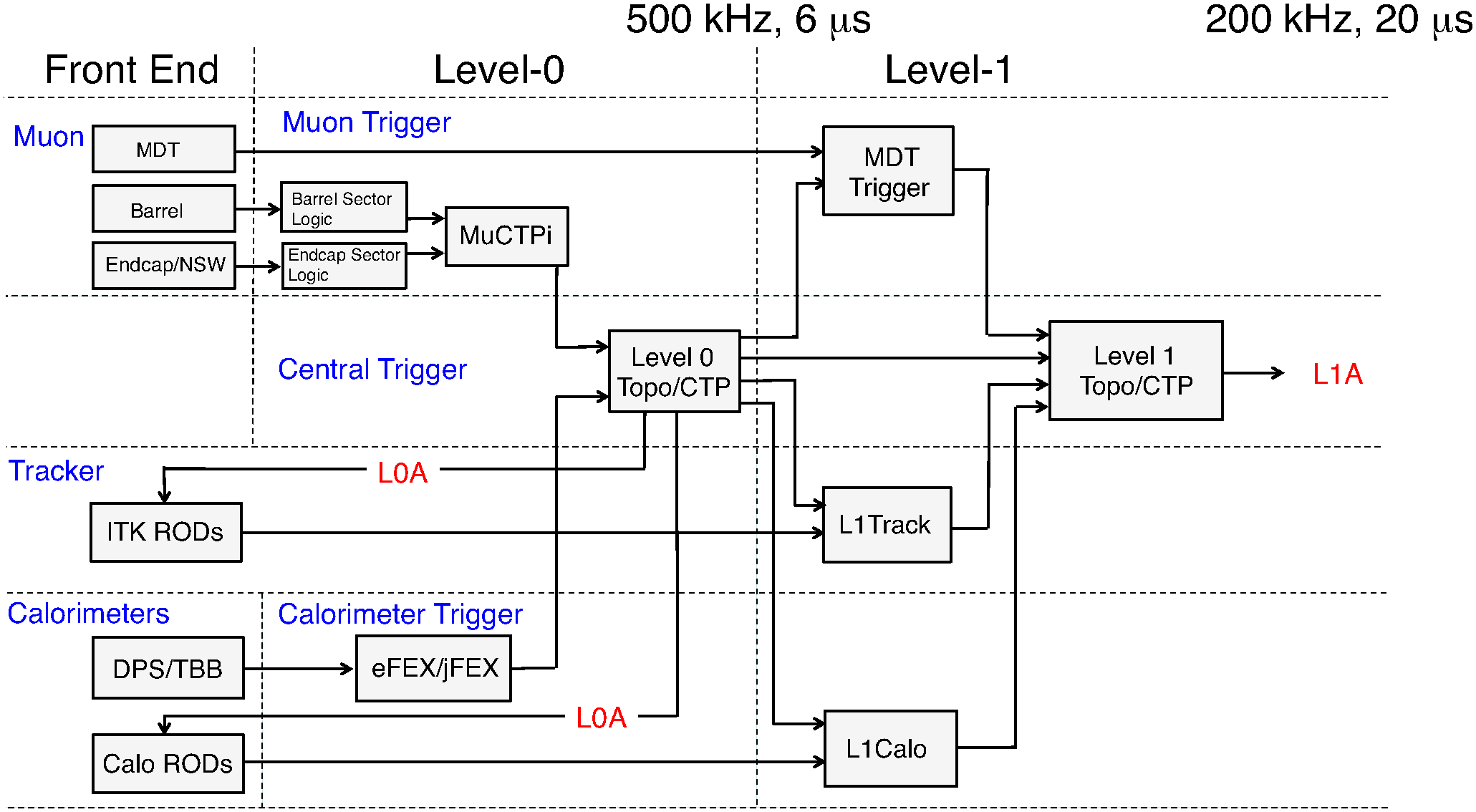} 
   \end{center}
   \vspace{-0.5cm}
\caption{A block diagram of the architecture of the split Level-0/Level-1 hardware trigger proposed for
the Phase-II upgrade.}
\label{fig:Trigger-II}
\end{figure}

 {\sc\emph{ \textbf{Trigger}}}.
A new trigger architecture is being developed that is compatible with the constraints imposed
by the detector and provides a flexible trigger with the potential to deliver the required performance.
As currently envisaged, the baseline design for the Phase-II Trigger, schematically sketched in Fig.~\ref{fig:Trigger-II}, foresees 
a split Level-0/Level-1 hardware trigger with a total level-1 accept rate of 200 kHz and
total latency of 20 $\mu$s.  The Level-0 trigger would distribute the Level-0 accept at a rate of at least 500 kHz within a
latency of 6 $\mu$s. The Phase-II Level-0 trigger is functionally the same as the Phase-I Level-1 system, based only on calorimetric and muon inputs. The Level-0 accept is generated by the
central trigger system which incorporates topological triggering capability.
The Level-1 system will reduce the rate to 200 kHz within an additional latency of 14 $\mu$s. This
reduction will be accomplished by the introduction of track information within a Region-of-Interest (RoI), full calorimeter granularity within the same RoI and the introduction of a
refined muon selection based on the use of the MDT information.
All the trigger electronics  of the calorimeters and muon spectrometer will need to be upgraded to face these trigger rates.  
Moreover the calorimeter electronics will also undergo changes to allow 40 MHz read-out of all the data without on-detector buffering.

\section{Conclusions}

A coherent overall upgrade program for ATLAS from Phase-0 through Phase-II has being planned to allow ATLAS to fully exploit the LHC energy and instantaneous luminosity at up to 5-7 times the design value.  Phase-0 is successfully concluded, while Technical Design Reports have been prepared for the Phase-I projects as well as the Letter of Intent for Phase-II. 
The planned approach is gradual and accommodates flexibility which can profit from running and indications from physics results.

\end{document}